\documentstyle [12pt]{article}
\begin{document}
\begin{center}
{\large\bf WAVELET ANALYSIS OF THE NEW SOLAR
NEUTRINO CAPTURE RATE DATA
FOR THE 
HOMESTAKE EXPERIMENT}
\end{center}
\vspace{1cm}
\begin{center}
{\bf H.J. Haubold} 
\end{center}
\noindent
Outer Space Office, United Nations, Vienna 
International Centre, P.O. Box 500, 1400 Vienna, Austria\par
\vspace{2cm}
\noindent
{\large Abstract.} Wavelet analysis offers an alternative to Fourier based time-series analysis and is
particularly useful when spectral features are time dependent. We analyze the solar
neutrino capture rate of the radiochemical Homestake chlorine experiment with abbreviated
Morlet wavelets, using Foster's (AJ, 111,1709(1996)) rescaled wavelet technique. We emphasize the complementarity of wavelet analysis to Fourier
analysis. Wavelet analysis confirms the results of previously undertaken Fourier analysis.
The Homestake data seem to contain a harmonic content with periodicities of 4.76 yr, 
1.89 yr, 0.85 yr, and 0.51 yr. Wavelet analysis reveals that the 4.76 yr and 1.89 yr periods
show an almost constant behavior over the 25 yr Homestake data record, while the 0.85 yr and
0.51 yr periods exhibit a transient phenomenon. The analysis does not show strong
evidence for a period of the solar 11 yr cycle.\par
\vspace{1cm}
\begin{tabbing}
1.1. \= \kill \\
1. \>The Solar Neutrino Problem\\
2. \>Is the Solar Neutrino Flux Constant Over Time ?\\
3. \> Wavelet Analysis of the Homestake Data Record\\
3.1.\> Wavelet and Fourier Spectra\\
3.2.\> Weighted Wavelet Z-Transform and Weighted Wavelet Amplitude\\
3.3.\> Results of the Wavelet Analysis\\
4. \>Conclusion\\
References\\
Table and Figure Captions
\end{tabbing}
\clearpage
\noindent
{\bf 1. The Solar Neutrino Problem}\par

Solar neutrinos were detected by four experiments, the radiochemical Homestake chlorine
experiment, the Kamiokande water Cherenkov experiment, and the two radiochemical
gallium experiments, GALLEX and SAGE. In these four experiments, typically less than
or of the order of 50 neutrino events were observed per year. These experiments confirmed
that the sun shines by nuclear fusion reactions among light elements, burning hydrogen
into helium, and that solar neutrinos have been observed in approximately the number and
with the energies expected (Bahcall 1989, 1996). In April, 1996, the Super-Kamiokande
experiment began to operate and inaugurated a new area of high-statistics tests of the
currently widely accepted standard solar models and standard electroweak theory (Suzuki
1997).\par
\medskip

The four experiments - Homestake, Kamiokande, GALLEX, and SAGE - have all
observed solar neutrino fluxes with intensities that are within a factor of a few of those
predicted by standard solar models. Three of the experiments (Homestake, GALLEX, and
SAGE) are radiochemical and each radiochemical experiment measures one number, the
total rate at which neutrinos above a fixed energy threshold, which depends on the
detector, are emitted by the sun. The only electronic (real-time, non-radiochemical)
detector among the four experiments, Kamiokande, has shown that the neutrinos come
from the sun, by measuring the recoil directions of the electrons scattered by solar
neutrinos (Bahcall 1996).\par
\medskip
 
Despite continual improvements of numerical standard solar model calculations of neutrino
fluxes over more than three decades, the discrepancies between the different solar neutrino
experiments and standard solar modeling have increased with time. All four of the solar
neutrino experiments yield neutrino capture rates that are significantly less than predicted
by standard solar models.\par
\medskip

Solar modeling is required in order to predict the rate of nuclear fusion by the proton-
proton chain reactions. In a standard solar model, about 99\% of the energy generation is
produced by reactions in the proton-proton chain. The most important neutrino producing
reactions are the low-energy pp (continuum: $0-0.4 MeV$), pep (one line: $1.4 MeV$), $^7Be$
(two lines: $0.86 MeV,\;\; 0.38 MeV$), and the high-energy $^8B$ (continuum: $0-14 MeV$)
neutrinos.\par
\medskip
In the simplest version of standard electroweak theory, neutrinos are massless and neutrino
flavors are separately conserved; neutrinos do not oscillate or decay to neutrinos with a
different lepton number or energy. The minimal standard electroweak theory has been
confirmed widely in precision laboratory tests (Barnett et al. 1996). Particle physics
generalizations of the standard electroweak theory (including left-right symmetry, grand
unification, and supersymmetry) suggest that the neutrino may have a mass in the range
$10^{-6} eV < m_\nu  < 10^{-2} eV$. Precision laboratory particle tests have not yet 
substantiated such
a suggestion.\par
\medskip

The first solar neutrino detection experiment to be performed was the crucial
radiochemical Homestake chlorine experiment, which detects electron neutrinos that are
more energetic than $0.81 MeV$. After almost three decades of operation of this experiment,
the measured neutrino capture rate is $2.56\pm0.22$ SNU, which is a factor $\sim 3.6$ less than is
predicted by the most detailed theoretical calculations, $9.3^{+1.2}_{-1.4}$ SNU (a SNU is a
convenient unit to describe the measured rates of solar neutrino experiments: $10^{-36}$
interactions per target atom per second). The predicted neutrino capture rate in the
Homestake experiment is dominated by the rare, high-energy $^8B$ neutrinos, although $^7Be$
neutrinos  also contribute significantly. According to standard model calculations, the pep
neutrinos and the CNO neutrinos are expected to contribute less than 1SNU to the total
neutrino capture rate. The discrepancy between the standard solar model calculations and
the observations of the radiochemical Homestake chlorine experiment came to be known
as the solar neutrino problem (Davis 1968, 1992, 1996; Zimmerman 1996).\par
\medskip

The comparison of the experimental results of the four solar neutrino experiments shows
(i) the smaller than predicted absolute event rates in the Homestake and Kamiokande
experiments, (ii) the incompatibility of the Homestake and Kamiokande experiments, and
(iii) the very low neutrino capture rates in the GALLEX and SAGE experiments which
may imply a great reduction of $^7Be$ neutrinos, although $^8$B neutrinos are observed, with
respect to the value predicted by the standard solar models. The conclusion based on these
three facts is that (i) either at least three of the four solar neutrino experiments (GALLEX
and SAGE and either Homestake and Kamiokande) have yielded misleading results, (ii)
physics beyond the standard electroweak theory is required to change the neutrino energy
spectrum or flavor content after the neutrinos are produced in the central region of the sun,
or (iii) the solar neutrino flux is varying over time due to unknown physical phenomena.
This paper is addressing the search for such a variation of the neutrino capture rate in the
Homestake experiment.\par
\medskip
\noindent
In Section 2 we review the current status of the search for periodicities in the Homestake
data record and summarize the results obtained by Fourier and Lomb-Scargle analysis of
the data.\par
\medskip
\noindent
The relationship between Fourier analysis and wavelet analysis is presented in Section 3.1.\par
\medskip
\noindent
The method of the weighted wavelet Z-transform and weighted wavelet amplitude is
outlined in Section 3.2. This special implementation of wavelet analysis takes into account
the fact that the Homestake data are unevenly spaced, contain spikes and gaps in the
record, and cover only a finite interval of time spanning 25 years.\par 
\medskip
\noindent
The results for the search for periodicities based on wavelet analysis are given in Section
3.3. Wavelet analysis confirms that the Homestake data contain a series of periodicities of
4.76 yr, 1.89 yr, 0.85 yr, and 0.51 yr. The time dependence of these spectral features
reveals that the first two periods exhibit an almost constant behavior over the 25 years of
Homestake data record, while the latter two periods show a transient nature with
variability of the actual value of the period. We find no strong evidence for a period of the
order of 11 yr in the data.\par
\medskip
\noindent
Conclusions are discussed briefly in Section 4.\par
\medskip
\noindent
{\bf 2. Is the Solar Neutrino Flux Constant Over Time ?}\par
\medskip
\noindent
The solar neutrino flux has been inferred from the neutrino capture rate in the Homestake
experiment (Figure 1), measured over the past 25 years (1970.281-1994.388) in 108 runs
(Davis 1968, 1992, 1996). Recently, the Homestake data have been completely reanalyzed
by the operators of the experiment, leading to changes in the uncertainties of the
measurements (Davis 1996). We emphasize this fact by dubbing them ``new'' solar neutrino
capture rate data. The observed neutrino capture rate is, on average, several times smaller
than the predicted one based on standard solar models. In the following we are not
concerned with the discrepancy between the average rate and the predicted rate, but with a
possible time dependence of the neutrino capture rate in the radiochemical Homestake
chlorine experiment. Suggestions that the neutrino capture rate is anticorrelated with solar
activity, reflected by the number of sunspots, were made by Davis (1992, 1996), in
particular when the neutrino capture rate went through an apparent minimum around 1980
(coinciding with the time when sunspot cycle 21 reached its maximum). Such a suggestion
was particularly encouraged when analyzing the five-month moving average of the
Homestake data record.\par 
\medskip

Bahcall, Field, and Press  (1987) and Bahcall and Press (1991) showed that conclusions
about correlation with sunspot number, essentially based on the Spearman rank order
correlation coefficient, Kendall's tau, and the ``shuffle test'',  change significantly when one
uses either the upper errors or the average errors of individual runs. In the following we
have chosen to limit the analysis of the solar neutrino capture rate in the radiochemical
Homestake chlorine experiment to the average value for each run, although the analysis of
Bahcall, Field, and Press  (1987) and Bahcall and Press (1991) indicated that the size of
the error bars for the runs with a very small neutrino capture rate is important for
determining the time dependence of the solar neutrino flux. Sturrock, Walther, and
Wheatland (1997) presented statistical tests for the variation of the neutrino capture rate in
the Homestake experiment and find evidence that the rate is not constant over time.
Particularly, they searched for a low-frequency content in the Homestake data and
concluded, through spectrum analysis based on the use of maximum likelihood estimation,
that there is no evidence for periods of the order of 11 yr (solar cycle) but some evidence
for periods of 2.1 yr (Sakurai periodicity; Sakurai 1979) and 0.44 yr (Rieger periodicity;
Rieger et al. 1984). We do emphasize here that at the current point of time the latter two
periodicities can not be related to or explained by any solar periodic process in the central
region of the sun based on standard solar modeling (Sakurai presented evidence that such
a quasi-biennial periodicity is contained in the Homestake data; Rieger et al. have shown
that the occurrence rate of solar flares exhibits a periodicity of about this length). Sturrock,
Walther, and Wheatland (1997) also searched the Homestake data for high-frequency
content, i.e. variations of the order of or less than the average exposure time of individual
runs, and find support for an earlier proposition (Sturrock and Bai 1992) that the neutrino
capture rate may be modulated at a frequency that could be related to the sidereal
rotational frequency of $13.88$ cycles yr$^{-1}$ of the sun's radiative zone.\par
\medskip

Standard solar modeling predicts that the neutrino fluxes are constant in time. Neither
solar physics nor nuclear physics of the standard solar model allow variation of the solar
neutrino flux over time. Additionally, the four solar neutrino experiments are low event
rates experiments with large error bars on each of the measured values. Despite of this, it
can not be excluded, a priori, that unknown physical phenomena may lead to a variation of
the flux additional to seasonal variations caused by the Earth's orbital eccentricity.\par 
\medskip

The simplest technique available for investigating periodicities in a data record is the
Fourier analysis; i.e., the comparison between a record and a sinusoidal signal with a
given frequency. Fourier analysis has a long tradition in analyzing ``quantities subject to
irregular fluctuations'' (Einstein 1914), however, the traditional Fourier transform method
often may not be suitable for representing the real structure of sparse data (Foster 1996a; Sturrock,
Walther, and Wheatland 1997). This shortcoming is evident when  suspected periods and
their amplitudes change with time and the sampling extends over a finite interval. The
application of Fourier analysis to such a record may lead to the identification of spurious
periodicities. Therefore, a method of avoiding the identification of  artificial periodicities is
needed, and wavelet transform is a technique that seems to suite this purpose (Strang
1993; Jawerth and Sweldens 1994). The wavelet method can reveal variations of spectral
features in a time dependent solar neutrino flux. Foster (1996b) introduced a rescaled
wavelet analysis technique that appears to be an effective method of reducing artificial
periodicities which may result from Fourier analysis. Three characteristics of the
Homestake record complicate period analysis. The first is the limited duration of the
record, the second is the presence of spikes and gaps in the record, and the third is the
uneven distribution of the data. Gaps are related to physical and practical constraints; i.e.,
the mean-half life of argon in the radiochemical Homestake chlorine experiment $(~35 days)$ and
facility maintenance. Standard wavelet methods require time series to be regularly
distributed in time. Extended gaps in the Homestake record can not be filled by
interpolation because of the unknown pattern of the signal itself. Commonly, these
constraints appear in the context of both Fourier and wavelet analysis. An extension of
Fourier technique that attempts to circumvent one of those limitations is the Lomb-Scargle
periodogram ( Lomb 1976; Scargle 1982). Its aim is to correct the trial functions $(cos\omega  t,
sin \omega  t)$ of the Fourier transform to preserve its normalization condition on an uneven grid
of times of a data record. For this purpose, an extraction of the mean value of the signal
under investigation and a phase shift of trial functions are used. The appealing feature of
the Lomb-Scargle normalized periodogram is that it weights data on a per-point rather than
a per-time-interval basis. Table 1 summarizes the results of previous Fourier and Lomb-Scargle
analysis of the Homestake data record, where the uneven distribution of the data had been
taken into account.\par
\medskip

In a way similar to that in which the Lomb-Scargle periodogram complements the Fourier
analysis, the wavelet analysis has been extended by Foster (1996b) to the rescaled wavelet
technique. The rescaled wavelet technique can be applied to short data series with spikes
and gaps. The idea behind this technique is basically the same as that behind the 
Lomb-Scargle periodogram: wavelet trial functions are extended and corrected by a rescale procedure to
satisfy the admissibility condition on the uneven grid of times of a data record. In this
technique the wavelet transform is understood as a projection onto trial functions which do
resemble the shape of the data record.\par
\medskip

We describe the rescaled wavelet technique and then compare the results obtained from
the rescaled wavelet transform technique to previous periodicity searches performed on the
Homestake record with Fourier analysis and  the Lomb-Scargle normalized periodogram
method (Table 1). Wavelet analysis allows the study of the variability of spectral features over time
unlike the Fourier transform.\par
\bigskip
\noindent
{\bf 3. Wavelet Analysis of the Homestake Data Record}\par
\medskip
\noindent
3.1. Wavelet and Fourier Spectra\par
\smallskip

Fourier analysis fails when one needs to consider the time dependence of spectral features.
By contrast, wavelets are efficient in multiscale analysis (Strang 1993; Jawerth and
Sweldens 1994). They have a localized, oscillating form so that, unlike sinusoids, they are
localized near time $\tau$  and decay if $|t-\tau|$  exceeds a characteristic scale, $a$     . Therefore,
the wavelet representation can be considered to be a mathematical microscope with
variable position and magnification. The wavelet transform represents one-dimensional
signals as a function of both time and frequency (position and scale) and is similar to a
local, filtered Fourier transform that can be obtained by dilating and translating the
wavelet and then convolving it with the signal (Walker 1997). 
We have a real signal $f$    of a real variable $t$  , represented by the data record 
$f(t)$. One
defines its Fourier transform $F(k)$ and its Fourier spectrum $P_F(k)$ accordingly,
\begin{equation}
F(k)=\int^{+\infty}_{-\infty}dt f(t)e^{-ikt},
\end{equation}
\begin{equation}
P_F(k)=\frac{1}{2\pi}|F(k)|^2 \;\;\;for\;\; k\geq0.
\end{equation}
\noindent
The total energy $E$ of the signal $f$  is defined such that
\begin{equation}
E=\frac{1}{2}\int^{+\infty}_{-\infty}dt|f(t)|^2=\frac{1}{4\pi}\int^{+\infty}_{-\infty}dk|F(k)|^2=\int^{+\infty}_0dkP_F(k).
\end{equation}
\noindent

The function $\psi$ is called an analyzing wavelet, if it verifies the admissibility condition
\begin{equation}
c_\psi=\int^{+\infty}_0 d\omega \omega^{-1}|\hat{\psi}(\omega)|^2\;\;\;< +\infty,
\end{equation}
\noindent
where
\begin{equation}
\hat{\psi}(\omega)=\int^{+\infty}_{-\infty}dt\psi(t)e^{-i\omega t}
\end{equation}
\noindent
is the Fourier transform of the wavelet. Condition (4) implies that the integrand defining    
$c_\psi$ should be integrable at $\omega=0$  and therefore that the wavelet has a zero mean, $\hat{\psi}(0)=0$.
The wavelet will also decay to zero as the variable tends to zero. A stronger condition is
to require cancellation up to some order p, that is
\begin{equation}
\int^{+\infty}_{-\infty}dt t^n \psi(t)=0 \;\;\;\mbox{for}\;\; n=0,1,\ldots,p-1,\;\;\mbox{and} \;\;\int^{+\infty}_{-\infty}dt t^p \psi(t)\neq 0.
\end{equation}
\noindent
If the function $\psi$ is integrable and square integrable, then the admissibility condition (4)
is equivalent to a cancellation condition of the order of at least zero and condition (6) is
equivalent, up to a multiplicative constant, to the existence of a bounded continuous
function $\phi$ with $\phi(0)=1$ and $\phi(\infty)=0$ such that
\begin{equation}
\hat{\psi}(\omega)=\omega^p\phi(\omega).
\end{equation}
\noindent
Additionally, the wavelet $\psi$ should be localized both in physical and in Fourier space
(time and frequency). This requirement means that the time spread, $\Delta t$, and the
frequency spread, $\Delta\omega$, of $\psi$ must satisfy the Heisenberg relation $\Delta t \Delta\omega = const.$,
where the constant cannot be smaller than $2\pi$. One defines the wavelet transform $W(a,\tau)$ of the signal $f$,
\begin{equation}
W(a, \tau)=\frac{1}{a^{1/2}}\int^{+\infty}_{-\infty}dt f(t)\psi^*(\frac{t-\tau}{a}),
\end{equation}
\noindent
where $a$ and $\tau$ denote the dilation (frequency) and translation (position) scaling factors,
respectively. For $|a|<<1$, the wavelet is a highly concentrated shrunken version of $\psi$     
with frequency content mostly in the high frequency range. If $|a|>>1$, the wavelet is
spread out and contains mainly low frequencies. In practice, the integral limits may be
replaced by values at which the wavelet amplitude is below some threshold, and $t$ is
evaluated at data points of the time series, which need not be the same times as $\tau$      . The
continuous wavelet transform can also be computed from the Fourier transform of the
signal $f$:
\begin{equation}
W(a,\tau)=\frac{a^{1/2}}{2\pi}\int^{+\infty}_{-\infty}d\omega F(\omega)\hat{\psi}^*(a\omega)e^{i\omega\tau}.
\end{equation}
\noindent
From the energy conservation
property of the wavelet transform,
\begin{equation}
E=\frac{1}{2c_\psi}\int^{+\infty}_0\frac{da}{a^2}\int^{+\infty}_{-\infty}d\tau\left|W(a,\tau)\right|^2,
\end{equation}
\noindent
we define the local wavelet spectrum
\begin{equation}
P_W(k,t)=\frac{1}{2c_\psi k_0}\left|W(\frac{k_0}{k},t)\right|^2\;\; \mbox{for}\;\; k \geq 0,
\end{equation}
\noindent
where $k_0$ denotes the peak frequency of the analyzing wavelet $\psi$. The local wavelet
spectrum measures the contribution to the total energy coming from the vicinity of the
point $t$ and frequency $k$, where the vicinity depends on the shape of the analyzing
wavelet in physical and Fourier space. From the local wavelet spectrum we derive a mean
wavelet spectrum $P_W(k)$:
\begin{equation}
P_W(k)=\int^{+\infty}_{-\infty}dt P_W(k,t),
\end{equation}
\noindent
which is directly related to the total energy $E$ of the signal $f$
\begin{equation}
E=\int^{+\infty}_0dk P_W(k).
\end{equation}
\noindent

Equations (2), (9), (11), and (12) reveal the relationship between the Fourier spectrum       
$P_F(\omega)$ and the mean wavelet spectrum  $P_W(k)$, namely,
\begin{equation}
P_W(k)=\frac{1}{c_\psi k}\int^{+\infty}_0 d\omega P_F(\omega)\left|\hat{\psi}(\frac{k_0\omega}{k})\right|^2.
\end{equation}
\noindent
Equation (14) shows that the mean wavelet spectrum is the average of the Fourier
spectrum weighted by the square of the Fourier transform of the analyzing wavelet $\psi$     
shifted at frequency $k$. The higher the frequency $k$ the wider the averaging interval.
Based on this behavior of the mean wavelet spectrum it can be shown that the mean
wavelet spectrum at high frequencies depends on the behavior of the analyzing wavelet at
low frequencies.\par

In the following we use the family of analyzing wavelets which are sinusoids in a
Gaussian envelope, known as modified Morlet wavelets. This family is well suited for
adjusting the tradeoff between time and frequency resolution,
\begin{equation}
\psi(z,n)=e^{-\frac{z^2}{2n^2\pi^2}}\left(e^{iz}-e^{-\frac{n^2\pi^2}{2}}\right)
\end{equation}
\noindent
where $z=\frac{t-\tau}{a}\;\;\mbox{and}\;\;\; e^{-\frac{n^2\pi^2}{2}}$                    is a small correction term which has been inserted
(modifying the Morlet wavelet) so that it satisfies the admissibility condition (4). The
parameter in the corrective term is related to Foster's (1996b) constant $c$ by $2n^2\pi^2=c^{-1}$,
which suits better the wavelet used in this analysis; it determines how rapidly the
analyzing wavelet decays. For $n \simeq 1$, the parameter $n$ corresponds to the number of
sinusoidal periods which fit between the inflection points of the Gaussian envelope. For     
$n\simeq 2$, the corrective term can even be neglected as it becomes very small. The larger the
value of $n$ the wider the window and the better the resolution in frequency but the worse
the resolution in time. In the following numerical wavelet analysis we will adopt the value 
$n=2\;\; (c\approx 0.0127)$, for this choice the corrective term is negligible and leads to the
abbreviated Morlet wavelet
\begin{equation}
\psi(z,n)=e^{iz-\frac{z^2}{2n^2\pi^2}}=e^{i\frac{(t-\tau)}{a}-\frac{(t-\tau)^2}{2n^2\pi^2a^2}}.
\end{equation}
\noindent
The chosen value $n=2$ fine-tunes the decay rate of the wavelet.\par

Employing the abbreviated Morlet wavelet leads to the abbreviated Morlet transform and,
for the discrete data record under consideration, it defines the discrete wavelet transform.
The abbreviated Morlet transform looks similar to a windowed Fourier transform, with the
window $e^{-\frac{(t-\tau)^2}{2n^2\pi^2a^2}}$, exhibiting the factor $a^{-2}$ in the exponential that makes the size of the window frequency dependent.
Assuming, for example, the total width of the abbreviated wavelet is about 7 years, we can
find the correlation between this curve and the first 7 years of the Homestake data record
shown in Figure 1. This single number gives a measure of the projection of this
wavelet on the data record during the 1970 - 1977 period; i.e. how much (amplitude) does
our 7 year period resemble a sine wave of this width (frequency). By sliding this wavelet
along the Homestake data record one can then construct a new data record of the
projection amplitude versus time. Eventually, one can vary the scale of the wavelet by
changing its width. This is the real advantage of wavelet analysis over a moving Fourier
spectrum. For a window of a certain width, the sliding Fourier transform is fitting different numbers of
waves; i.e. there can be many high-frequency waves within a window, while the same
window can only contain a few (or even less than one) low-frequency waves. The wavelet
analysis uses a wavelet of the exact same shape, only the size scales up and down
with the size of the window.\par
\bigskip
\noindent
3.2. Weighted Wavelet Z-Transform and Weighted Wavelet Amplitude\par
\smallskip
\noindent
The discrete analogue to (8) is
\begin{equation}
W(a,\tau)=\frac{1}{a^{1/2}}\sum^N_{\alpha =1}f(t_\alpha)\psi^*\left(\frac{t_\alpha-\tau}{a}\right)
\end{equation}
for the observed time series consisting of N data values $f(t_\alpha)$, taken at a discrete set of N times $t_\alpha, \alpha = 1,2,...,N$. For the abbreviated Morlet wavelet (16) the real and imaginary parts of the discrete wavelet transform are
\begin{equation}
Re(W)=\frac{1}{a^{1/2}}\sum^N_{\alpha=1}f(t_\alpha)e^{-\frac{(t_\alpha-\tau)^2}{2n^2\pi^2a^2}}cos \left(\frac{t_\alpha-\tau}{a}\right)
\end{equation}
and
\begin{equation}
Im(W)=-\frac{1}{a^{1/2}}\sum^N_{\alpha=1}f(t_\alpha)e^{-\frac{(t_\alpha-\tau)^2}{2n^2\pi^2a^2}}sin \left(\frac{t_\alpha-\tau}{a}\right).
\end{equation}
When treating the discrete wavelet transform (17) as a projection, it can be interpreted as a weighted projection onto the trial function
\begin{equation}
\Phi(t) = e^{i\frac{t-\tau}{a}}
\end{equation}
with the statistical weights chosen as
\begin{equation}
w_\alpha=e^{-\frac{(t_\alpha-\tau)^2}{2n^2\pi^2a^2}}.
\end{equation}
In order to tackle the problem of uneven time spacing with the discrete wavelet transform, similar to a procedure used in discrete Fourier tansform, Foster (1996b) suggested to include a third trial function, the constant function ${\bf 1}(t)=1$ for all t. Therefore, a weighted projection onto three trial functions is performed, namely
\begin{eqnarray}
\Phi_1(t)& = & {\bf 1}(t),\label{line1}\\
\Phi_2(t)& = & cos(\frac{t-\tau}{a}),\label{line2}\\
\Phi_3(t) & = &sin (\frac{t-\tau}{a}).\label{line3}
\end{eqnarray}
Projection determines the coefficient $y_a$ of a set of $r$ trial functions $\Phi_a(t), a= 1,2,\ldots, r$ for which the function under consideration,
\begin{equation}
y(t)=\sum_a y_a(t)\Phi_a(t),
\end{equation}
fits best the data record. Such a projection can be computed by defining the inner product of two functions $u(t)$ and $v(t)$ as
\begin{equation}
<u|v>=\frac{\sum^N_{\alpha=1}w_\alpha u(t_\alpha)v(t_\alpha)}{\sum^N_{\beta=1}w_\beta},
\end{equation}
where $w_\alpha$ denotes the statistical weight assigned to the data point $\alpha$. The best-fit coefficients of the trial functions are determined by multiplying the inverse of the S-matrix, that is the matrix of inner products (26) of the trial functions $(S_{ab}=<\Phi_a|\Phi_b>)$ by the vector of inner products of the trial functions with the data record
\begin{equation}
y_a=\sum_b S_{ab}^{-1}<\Phi_b|f>.
\end{equation}
Based on the above procedure, Foster (1996a) defines the power for evaluating the projection statistically as
\begin{equation}
P=\frac{N}{(r-1)s^2}\left(\sum_{a,b}S_{a,b}^{-1}<\Phi_a|f><\Phi_b|f>-<{\bf 1}| f>^2\right)
\end{equation}
for the set of trial functions, where N is the number and $s^2$ the estimated variance of the data record. The power (28) is a chi-square statistic with $r-1$ degrees of freedom and expected value 1.

In the case of a weighted projection, taking into account (20) and (21), the number of data N in (28) has to be replaced by an effective number (Foster 1996a, 1996b)
\begin{equation}
N_{eff}=\frac{(\sum w_\alpha)^2}{(\sum w_\alpha^2)}=\frac{\left[\sum e^{-\frac{(t-\tau)^2}{2n^2\pi^2a^2}}\right]^2}{\sum e^{-\frac{(t-\tau)^2}{n^2\pi^2a^2}}},
\end{equation}
and for the variance $s^2$ in (28) the weighted estimated variance
\begin{equation}
s_w^2=\frac{N_{eff}V_f}{N_{eff}-1}
\end{equation}
has to be used. Accordingly, in (30) $V_f$ denotes the weighted variation of the data record,
\begin{equation}
V_f=\frac{\sum_\alpha w_\alpha f^2(t_\alpha)}{\sum_\lambda w_\lambda}-\left[\frac{\sum_\alpha w_\alpha f(t_\alpha)}{\sum_\lambda w_\lambda}\right]^2=<f|f>-<{\bf 1}|f>^2.
\end{equation}
and $V_y$ is the weighted variation of the function $y$
\begin{equation}
V_y=\frac{\sum_\alpha w_\alpha y^2(t_\alpha)}{\sum_\lambda w_\lambda}-\left[\frac{\sum_\alpha w_\alpha y(t_\alpha)}{\sum_\lambda w_\lambda}\right]^2=<y|y>-<{\bf 1}|y>^2.
\end{equation}
Thus, Foster's (1996b) weighted wavelet transform (WWT) is defined by 
\begin{equation}
WWT=\frac{(N_{eff}-1)V_y}{2V_f},
\end{equation}
for fixed $a$ and $\tau$, as a chi-square statistic with two degrees of freedom and expected value one.

The weighted wavelet transform in equation (33) depends sensitively on the effective
number of data $N_{eff}$ which leads to a shift of the WWT peaks to lower frequencies
because at lower frequencies the window is wider and thus more data points are sampled
and the effective number $N_{eff}$ is larger. To compensate for this fact, a test statistics
which is less sensitive to the effective number of data can be applied. Such a statistics for
projections was suggested by Foster (1996a), which leads to the weighted wavelet Z-
transform (WWZ)
\begin{equation}
WWZ=\frac{(N_{eff} -3)V_y}{2(V_f-V_y)}
\end{equation}
The statistical behavior of (34) is derived for a projection where the statistical weights are
inversely proportional to the variance of the data.\par

Although the WWZ may be an excellent locator of the signal frequency, it is a poor
measure of amplitude. With a projection it is easy to define the amplitude of the
corresponding periodic fluctuation: it is the square root of the sum of the squares of the
expansion coefficients for the sine and cosine functions, which Foster (1996b) calls the
weighted wavelet amplitude (WWA),
\begin{equation}
WWA=((y_2)^2+(y_3)^2)^{1/2}.
\end{equation}
Therefore the weighted wavelet transform allows us to compute the amplitude by the
WWA, after determining the period from the WWZ.\par
\medskip
\noindent
3.3. Results of the Wavelet Analysis\par
\smallskip

For the Homestake data we have subtracted the average value from the data prior to analysis.
The WWZ was computed over the frequency range 0 - 2.5 cycles yr$^{-1}$ in steps of 0.01
cycles yr$^{-1}$. It must be noted that we do read off the
significance of a peak directly from the heights of the peaks in the WWZ. Fig. 2 shows the perspective wavelet map for the analysed Homestake data. \par

In examining the WWZ, as seen in Fig. 3, in the neighborhood of the frequency of the solar
cycle (0.09 cycles yr$^{-1}$) we find no strong evidence of a peak, even if the WWZ shows a peak at 0.12 cycles yr$^{-1}$. This is compatible with
the analysis of Haubold and Gerth (1990), Haubold and Beer (1991), and Sturrock,
Walther, and Wheatland (1997). The examination of the spectrum in the neighborhood of
the Sakurai quasi-biennial periodicity (0.45 - 0.50 cycles yr$^{-1}$) and the Rieger periodicity 
(supposed to be located at 2.30 - 2.40 cycles yr$^{-1}$) we find there are peaks with high power at frequencies 0.53 and
1.93 cycles yr$^{-1}$, respectively. However, the power spectrum reveals even stronger peaks
at frequencies 0.21 cycles yr$^{-1}$ and 1.18 cycles yr$^{-1}$ which can not be related to any
known solar periodic process in the central region of the sun at this point of time. The most prominent peaks in the WWZ
are the frequencies at 0.21 and 0.53 cycles yr$^{-1}$.
Inspection of the development of the WWZ in Fig. 4 over the 25 yr of Homestake record indicates
that the value of the frequencies at 0.21 cycles yr$^{-1}$ and 0.53 cycles yr$^{-1}$ are stable at
their positions in the WWZ, showing, however, strong variation in the value of the WWZ.
Contrary, the frequencies 1.18 cycles yr$^{-1}$ and 1.93 cycles yr$^{-1}$ exhibit a wide variation
of their values in the WWZ over 25 yr of Homestake record, pointing towards a
transient phenomenon. 
From inspection of Figures 5 and 6, showing the Fourier spectrum  and amplitude of the
Homestake data, respectively, computed over the same frequency range with the same
frequency step as the WWZ, we find evidence for peaks near 0.21 cycles yr$^{-1}$ and 0.53
cycles yr$^{-1}$ as well as near 1.18 cycles yr$^{-1}$ and 1.93 cycles yr$^{-1}$ which have been
also revealed in the wavelet analysis.\par

Examination of the WWZ and power spectra shows that a series of periodicities (0.21,
0.53, 1.18, 1.93 cycles yr$^{-1}$) shows up in each analysis.\par
\bigskip
\noindent
{\bf 4. Conclusion}\par
\medskip

In a way similar to that in which Lomb-Scargle periodogram complements Fourier analysis,
Foster's weighted wavelet Z-transform extends wavelet analysis. In addition, the WWZ
allows to study the variability of periodicities over time, unlike Fourier analysis. The
computation of the WWZ confirms, to a reasonable extend, the results of Lomb-Scargle
and Fourier analysis, summarized in Table 1, applied to the unevenly spaced data yielded by the radiochemical
Homestake chlorine experiment, in that the data have a harmonic content. Analysis, using
the abbreviated Morlet wavelet, indicates that this harmonic content is represented by a
series of periodicities of 0.21, 0.53, 1.18, and 1.93 cycles yr$^{-1}$.\\
The two ``low-frequency'' values of them (0.21 and 0.53 cycles yr$^{-1}$) show a stable
behavior over the analyzed 25 years of Homestake record while the latter two frequencies
exhibit a transient behavior by large variation of their values over the 25 yr of data
analyzed. Two periodicities correspond with Sakurai's periodicity (0.45 - 0.50 cycles yr$^{-
1}$) and Rieger's periodicity (2.30 - 2.40 cycles yr$^{-1}$). None of the above analysis shows
strong evidence for a variation in the Homestake data with the 11 yr solar cycle.\par
\smallskip

A variation of the solar neutrino capture rate in the Homestake experiment may result
from fluctuations of nuclear fusion or by a modulation of the capture rate due to the
solar magnetic field or helioseismic waves but both phenomena are not supported by
standard solar modeling (Sturrock, Walther, and Wheatland 1997).\par
\bigskip 
\noindent
Acknowledgements. The author acknowledges gratefully information concerning the Homestake experiment provided by Raymond Davis Jr. \\
Wavelet analysis was performed using the compiler programm WWZ, developed by the American Association of Variable Star Observers; in this regard the co-operation of Grant Foster is appreciated.
Initial results of this work where presented and discussed at the Workshop on Data Analysis Techniques (November 1997) at the Brazilian Institute for Space Research (INPE); the author thanks the local organizers for the hospitality and stimulating atmosphere at Sao Jose dos Campos.
I am grateful to many colleagues for helpful discussions, particularly during the Fourth International Solar Neutrino Conference (April 1997) in Heidelberg, Germany.\par
\clearpage
\noindent
{\bf References}\par
\medskip
\noindent
Bahcall, J.N. 1989, Neutrino Astrophysics (Cambridge: Cambridge\par
University Press)\par
\smallskip
\noindent
Bahcall, J.N. 1996, ApJ, 467, 475\par
\smallskip
\noindent
Bahcall, J.N., Field, G.B., and Press, W.H. 1987, ApJ, 320, L69\par
\smallskip
\noindent
Bahcall, J.N., and Press, W.H. 1991, ApJ, 370, 730\par
\smallskip
\noindent
Barnett, R.M. et al. 1996, Phys. Rev., D54, 1\par
\smallskip
\noindent
Davis Jr., R. 1992, Ann. N.Y. Acad. Sci., 655, 209\par
\smallskip
\noindent
Davis Jr., R. 1996, Nucl. Phys., B48, 284\par
\smallskip
\noindent
Davis Jr., R., Harmer, D.S., and Hoffman, K.C. 1968, Phys. Rev. Lett., 20,\par 
1205\par
\smallskip
\noindent
Einstein, A. 1914, Archives de Sciences Physiques et Naturalles, 37, 254\par 
(English translation: IEEE ASSP Magazine, October 1987, 6)\par
\smallskip
\noindent
Foster, G. 1996a, AJ, 111, 541\par
\smallskip
\noindent
Foster, G. 1996b, AJ, 112, 1709\par
\smallskip
\noindent
Haubold, H.J., and Gerth, E. 1990, Sol. Phys., 127, 347\par
\smallskip
\noindent
Haubold, H.J., and Beer, J. 1991, in Solar-Terrestrial Variability and Global\par 
Change, eds. W. Schroeder and J.P. Legrand (Vienna: IUGG/IAGA\par 
General Assembly), 11\par
\smallskip
\noindent
Jawerth, B., and Sweldens, W. 1994, SIAM Review, 36, 377\par
\smallskip
\noindent
Lomb, N.R. 1976, A\&SS, 39, 447\par
\smallskip
\noindent
Rieger, E., et al. 1984, Nature, 312, 623\par
\smallskip
\noindent
Sakurai, K. 1979, Nature, 278, 146\par
\smallskip
\noindent
Scargle, J.D. 1982, ApJ, 263, 835\par
\smallskip
\noindent
Sturrock, P.A., and Bai, T. 1992, ApJ, 397, 337\par
\smallskip
\noindent
Sturrock, P.A., Walther, G., and Wheatland, M.S. 1997, ApJ, 491, 409\par
\smallskip
\noindent
Strang, G. 1993, Bull. (New Series) Amer. Math. Soc., 28, 288\par
\smallskip
\noindent
Suzuki, Y. 1997, in Fourth International Solar Neutino Conference, ed.\par 
W. Hampel (Heidelberg: Max-Planck-Institute for Nuclear Physics), 163\par
\smallskip
\noindent
Walker, J.S.. 1997, Not. Amer. Math. Soc., 44, 658\par
\smallskip
\noindent
Zimmerman, R. 1996, The Sciences (N.Y. Acad. Sci.), 36(1), 16\par
\clearpage
\noindent
{\bf Table and Figure Captions}\par
\medskip
\noindent
Table 1. Summary of periods $\pi$ revealed by Fourier, Lomb-Scargle, and wavelet analysis
of the Homestake record as discussed in the text and in Haubold and Gerth (1990) and
Haubold and Beer (1991).\par
\medskip
\noindent
Fig. 1. A plot of the individual solar neutrino capture rates from the Homestake
experiment, runs nos. 18 to 133 (1970.281 - 1994.388) (Davis 1996).\par
\medskip
\noindent
Fig. 2. The perspective wavelet map (frequency-time-WWZ) for the Homestake record.
The value of WWZ is approximately an F-statistic with N$_{eff}$ (the effective number of
data for the given time and frequency being analyzed; eq (29)), two degrees of freedom,
and expected value one. The graph indicates whether or not there is a periodicity at a
given time of the given frequency.\par
\medskip
\noindent
Fig. 3. The value of WWZ from Fig. 2 as a function of frequency showing the series of
periodicities in the Homestake record as discussed in the text.\par
\medskip
\noindent
Fig. 4. The value of the WWZ from Fig. 2 showing the time evolution of the amplitude of
the periodicities in the Homestake record.\par
\medskip
\noindent
Fig. 5. Spectrum of the Homestake data for low-frequencies, obtained by Fourier analysis.\par
\medskip
\noindent
Fig. 6. Amplitude of the Homestake data, as determined by Fourier analysis, corresponding
to the spectrum shown in Fig. 5.\par
\begin{tabular}{|lr|c|c|c|c|}\\ \hline
&$\pi[yr]$ & & & &  \\
&& $\pi<1$ &$1\leq\pi<2$ & $2\leq\pi<3$ & $3\leq\pi<4$ \\ 
 No.runs & & & & &   \\ \hline
 Fourier& & & & &   \\
 18-69 && 0.7 & 1.63 & 2.14 & 3.00\\ 
(1970-1981)& & 0.5 & 1.30 & &  \\ \hline
 Fourier && 0.83 & 1.61 & 2.13 & \\
 18-89 && 0.61 & & &\\
 (1970-1985)& & 0.54 & & & \\
 && 0.51 & & &  \\ \hline
 Lomb-Scargle & & 0.7 & 1.30 & & \\ 
 18-109& & 0.54 & & &   \\
(1970-1990)& & & & & \\ \hline
 Fourier && 0.7 & 1.75 & 2.04 & \\
 18-133& & 0.55 & 1.59 & & \\
 (1970-1994)& & 0.53 & & & \\
&&0.51 & & &  \\ \hline
 Wavelet& & 0.85 & 1.89 & & \\
 18-133 && 0.51 & & &  \\
(1970-1994)& & & & & \\ \hline
&$\pi[yr]$ & & & &  \\
&& $4\leq\pi<5$ & $5\leq\pi<6$ & $8\leq\pi<9$ & $9\leq\pi<10$ \\
No.runs & & & & & \\ \hline
Fourier& & & & 8.33 &  \\
18-69 & & 4.90 & & &\\
(1970-1981)& & & & & \\ \hline
Fourier & & & & 8.33 & \\
18-89 & & & & & \\
(1970-1985) & & & & & \\
& & & & &  \\ \hline
Lomb-Scargle &  & 4.80 & & & 9.60 \\
18-109 & & & & & \\
(1970-1990) & & & & & \\ \hline
Fourier &  &  4.55 & & & \\
18-133 & & & & & \\
(1970-1994) & & & & & \\
& & & & & \\ \hline
Wavelet &  & 4.76 & & & \\
18-133 & & & & & \\
(1970-1994) & & & & & \\\hline
\end{tabular}\par
\vspace*{1cm}
Table 1.
\thispagestyle{empty}
\end{document}